\documentclass[twoside]{article}
\usepackage{graphicx}
\usepackage{amsmath}

\catcode`\@=11
\long\def\@makefntext#1{
\protect\noindent \hbox to 3.2pt {\hskip-.9pt
$^{{\eightrm\@thefnmark}}$\hfil}#1\hfill}		

\def\@makefnmark{\hbox to 0pt{$^{\@thefnmark}$\hss}}	

\def\ps@myheadings{\let\@mkboth\@gobbletwo
\def\@oddhead{\hbox{}
\rightmark\hfil\eightrm\thepage}
\def\@oddfoot{}\def\@evenhead{\eightrm\thepage\hfil
\leftmark\hbox{}}\def\@evenfoot{}
\def\sectionmark##1{}\def\subsectionmark##1{}}

\oddsidemargin=\evensidemargin
\newcounter{sectionc}\newcounter{subsectionc}\newcounter{subsubsectionc}
\renewcommand{\section}[1] {\vspace{12pt}\addtocounter{sectionc}{1}
\setcounter{subsectionc}{0}\setcounter{subsubsectionc}{0}\noindent
	{\tenbf\thesectionc. #1}\par\vspace{5pt}}
\renewcommand{\subsection}[1] {\vspace{12pt}\addtocounter{subsectionc}{1}
	\setcounter{subsubsectionc}{0}\noindent
	{\bf\thesectionc.\thesubsectionc. {\kern1pt \bfit #1}}\par\vspace{5pt}}
\renewcommand{\subsubsection}[1] {\vspace{12pt}\addtocounter{subsubsectionc}{1}
	\noindent{\tenrm\thesectionc.\thesubsectionc.\thesubsubsectionc.
	{\kern1pt \tenit #1}}\par\vspace{5pt}}
\newcommand{\nonumsection}[1] {\vspace{12pt}\noindent{\tenbf #1}
	\par\vspace{5pt}}

\newcounter{appendixc}
\newcounter{subappendixc}[appendixc]
\newcounter{subsubappendixc}[subappendixc]
\renewcommand{\thesubappendixc}{\Alph{appendixc}.\arabic{subappendixc}}
\renewcommand{\thesubsubappendixc}
	{\Alph{appendixc}.\arabic{subappendixc}.\arabic{subsubappendixc}}

\renewcommand{\appendix}[1] {\vspace{12pt}
        \refstepcounter{appendixc}
        \setcounter{figure}{0}
        \setcounter{table}{0}
        \setcounter{lemma}{0}
        \setcounter{theorem}{0}
        \setcounter{corollary}{0}
        \setcounter{definition}{0}
        \setcounter{equation}{0}
        \renewcommand{\thefigure}{\Alph{appendixc}.\arabic{figure}}
        \renewcommand{\thetable}{\Alph{appendixc}.\arabic{table}}
        \renewcommand{\theappendixc}{\Alph{appendixc}}
        \renewcommand{\thelemma}{\Alph{appendixc}.\arabic{lemma}}
        \renewcommand{\thetheorem}{\Alph{appendixc}.\arabic{theorem}}
        \renewcommand{\thedefinition}{\Alph{appendixc}.\arabic{definition}}
        \renewcommand{\thecorollary}{\Alph{appendixc}.\arabic{corollary}}
        \renewcommand{\theequation}{\Alph{appendixc}.\arabic{equation}}
        \noindent{\tenbf Appendix \theappendixc #1}\par\vspace{5pt}}
\newcommand{\subappendix}[1] {\vspace{12pt}
        \refstepcounter{subappendixc}
        \noindent{\bf Appendix \thesubappendixc. {\kern1pt \bfit #1}}
	\par\vspace{5pt}}
\newcommand{\subsubappendix}[1] {\vspace{12pt}
        \refstepcounter{subsubappendixc}
        \noindent{\rm Appendix \thesubsubappendixc. {\kern1pt \tenit #1}}
	\par\vspace{5pt}}

\newcommand{\textlineskip}{\baselineskip=13pt}
\newcommand{\smalllineskip}{\baselineskip=10pt}

\def\eightcirc{
\begin{picture}(0,0)
\put(4.4,1.8){\circle{6.5}}
\end{picture}}
\def\eightcopyright{\eightcirc\kern2.7pt\hbox{\eightrm c}}

\newcommand{\copyrightheading}[1]
	{\vspace*{-2.5cm}\smalllineskip{\flushleft
	{\footnotesize International Journal of Modern Physics C, #1}\\
	{\footnotesize $\eightcopyright$\, World Scientific Publishing
	 Company}\\
	 }}

\newcommand{\publisher}[2]{{\begin{center}\footnotesize\smalllineskip
	Received #1\\
	Revised #2
	\end{center}
	}}

\def\abstracts#1#2#3{{
	\centering{\begin{minipage}{4.5in}\baselineskip=10pt\footnotesize
	\parindent=0pt #1\par
	\parindent=15pt #2\par
	\parindent=15pt #3
	\end{minipage}}\par}}

\def\keywords#1{{
	\centering{\begin{minipage}{4.5in}\baselineskip=10pt\footnotesize
	{\footnotesize\it Keywords}\/: #1
	\end{minipage}}\par}}

\renewenvironment{thebibliography}[1]
        {\frenchspacing
	 \ninerm\baselineskip=11pt
         \begin{list}{\arabic{enumi}.}
        {\usecounter{enumi}\setlength{\parsep}{0pt}
	 \setlength{\leftmargin 12.7pt}{\rightmargin 0pt} 
         \setlength{\itemsep}{0pt} \settowidth
	{\labelwidth}{#1.}\sloppy}}{\end{list}}

\newcounter{itemlistc}
\newcounter{romanlistc}
\newcounter{alphlistc}
\newcounter{arabiclistc}

\newcommand{\fcaption}[1]{
        \refstepcounter{figure}
        \setbox\@tempboxa = \hbox{\footnotesize Figure~\thefigure. #1}
        \ifdim \wd\@tempboxa > 5in
           {\begin{center}
        \parbox{5in}{\footnotesize\smalllineskip Figure~\thefigure. #1}
            \end{center}}
        \else
             {\begin{center}
             {\footnotesize Figure~\thefigure. #1}
              \end{center}}
        \fi}

\newcommand{\tcaption}[1]{
        \refstepcounter{table}
        \setbox\@tempboxa = \hbox{\footnotesize Table~\thetable. #1}
        \ifdim \wd\@tempboxa > 5in
           {\begin{center}
        \parbox{5in}{\footnotesize\smalllineskip Table~\thetable. #1}
            \end{center}}
        \else
             {\begin{center}
             {\footnotesize Table~\thetable. #1}
              \end{center}}
        \fi}

\def\@citex[#1]#2{\if@filesw\immediate\write\@auxout
	{\string\citation{#2}}\fi
\def\@citea{}\@cite{\@for\@citeb:=#2\do
	{\@citea\def\@citea{,}\@ifundefined
	{b@\@citeb}{{\bf ?}\@warning
	{Citation `\@citeb' on page \thepage \space undefined}}
	{\csname b@\@citeb\endcsname}}}{#1}}

\newif\if@cghi
\def\cite{\@cghitrue\@ifnextchar [{\@tempswatrue
	\@citex}{\@tempswafalse\@citex[]}}
\def\citelow{\@cghifalse\@ifnextchar [{\@tempswatrue
	\@citex}{\@tempswafalse\@citex[]}}
\def\@cite#1#2{{$\null^{#1}$\if@tempswa\typeout
	{IJCGA warning: optional citation argument
	ignored: `#2'} \fi}}
\newcommand{\citeup}{\cite}

\def\pmb#1{\setbox0=\hbox{#1}
	\kern-.025em\copy0\kern-\wd0
	\kern.05em\copy0\kern-\wd0
	\kern-.025em\raise.0433em\box0}

\def\fnt#1#2{\footnotetext{\kern-.3em
	{$^{\mbox{\scriptsize #1}}$}{#2}}}

\def\fpage#1{\begingroup
\voffset=.3in
\thispagestyle{empty}\begin{table}[b]\centerline{\footnotesize #1}
	\end{table}\endgroup}

\def\runninghead#1#2{\pagestyle{myheadings}
\markboth{{\protect\footnotesize\it{\quad #1}}\hfill}
{\hfill{\protect\footnotesize\it{#2\quad}}}}
\font\tenrm=cmr10
\font\tenit=cmti10
\font\tenbf=cmbx10
\font\bfit=cmbxti10 at 10pt
\font\ninerm=cmr9

\font\eightrm=cmr8

\def\qed{\hbox{${\vcenter{\vbox{			
   \hrule height 0.4pt\hbox{\vrule width 0.4pt height 6pt
   \kern5pt\vrule width 0.4pt}\hrule height 0.4pt}}}$}}

\def\bsc{{\sc a\kern-6.4pt\sc a\kern-6.4pt\sc a}}  	
\def\bflatex{\bf L\kern-.30em\raise.3ex\hbox{\bsc}\kern-.14em 
T\kern-.1667em\lower.7ex\hbox{E}\kern-.125em X}

\begin{document}
\runninghead{K.~Malarz \& A.~Z.~Maksymowicz}{A simple solid-on-solid model of epitaxial films growth: surface morphology anisotropy}
\normalsize\textlineskip
\thispagestyle{empty}
\setcounter{page}{1}
\copyrightheading {Vol. 10, No. 0 (1999) 000--000}
\vspace*{0.88truein}
\fpage{1}
\centerline{\bf A SIMPLE SOLID-ON-SOLID MODEL}
\centerline{\bf OF EPITAXIAL FILM GROWTH:}
\centerline{\bf SURFACE MORPHOLOGY ANISOTROPY}
\vspace*{0.37truein}
\centerline{\footnotesize K.~MALARZ$^*$ and A.~Z.~MAKSYMOWICZ$^\dag$}
\vspace*{0.015truein}
\centerline{\footnotesize\it Department of Theoretical and Computational Physics,}
\centerline{\footnotesize\it Faculty of Physics and Nuclear Techniques, University of Mining and Metallurgy (AGH)}
\centerline{\footnotesize\it al. Mickiewicza 30, PL-30059 Krak\'ow, Poland}
\centerline{\footnotesize\it E-mail: $^*${\tt malarz@agh.edu.pl}, $^\dag${\tt amax@agh.edu.pl}}
\vspace*{0.225truein}
\publisher{(received date)}{(revised date)}
\vspace*{0.21truein}
\abstracts{In this paper we present a generalization of a simple solid-on-solid epitaxial model of thin films growth, when surface morphology anisotropy is provoked by anisotropy in model control parameters: binding energy and/or diffusion barrier.
The anisotropy is discussed in terms of the height-height correlation function.
It was experimentally confirmed that the difference in diffusion barriers yields anisotropy in morphology of the surface.
We got antisymmetric correlations in the two in-plane directions for antisymmetric binding.}{}{}
\vspace*{10pt}
\keywords{Surface structure, morphology, roughness, and topography; Surface diffusion; Computer simulations, Monte Carlo methods}
\vspace*{1pt}\textlineskip

\section{Introduction}
\vspace*{-0.5pt}
\noindent
The surface growth phenomena happen very often, not only in physics.
It is the case of crystal growth,\citeup{levi97,kotrla92a} epidemics, live cell growth, corrosion of materials, the spreading of forest fires, {\it etc}.\citeup{herrmann86,gouyet91}
Anisotropic growth case could be considered for example of the fire as result of influence of strong wind in one direction, or crystals growing preferentially by diffusion in an easy diffusion, {\it etc}.
In this paper we would like to present some suitable modifications of our earlier model\cite{maksymowicz96}\cite{malarz99a} for the anisotropic growth case when model control parameters are different in different directions.

	The rules of growth are based on deposition model with particles diffusion limited to one-step, to the nearest neighbor only.\cite{malarz99a}
The probability of a particle settling at a position is given by the Boltzmann factor $\exp(-E/kT)$ where $E$ is particle energy at the position, $k$ is the Boltzmann constant, and $T$ denotes temperature.
After this one inter-atomic-separation long step particle sticks there and stays for rest of simulation.
We consider three dominant contributions to the particle energy: binding to underlying atom energy $S\le 0$, nearest neighbor interaction energy $J$, and diffusion barrier $V\ge 0$.
We assume that the model control parameters may be different in different directions ($S_x\ne S_y$ or $J_x\ne J_y$ or $V_x\ne V_y$), which is essential to generate the claimed surface morphology anisotropy.

\begin{figure}
\centering
\includegraphics[height=35mm]{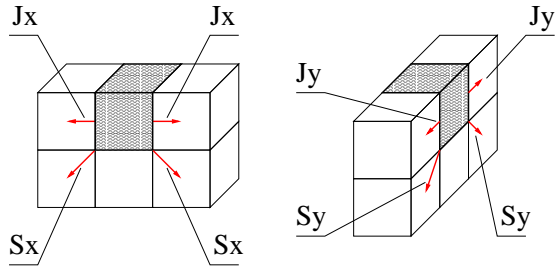}
\caption{Model control parameters $S$ and $J$.
Indexes $x$ and $y$ correspond to $x$- or $y$-directions of atom pairs.
Anisotropy results from $J_x\ne J_y$ or $S_x\ne S_y$.}
\label{model}
\end{figure}

	The nearest neighbor energy $J$ acting between neighbors on same film layer, had to be enriched by the next nearest neighbor interaction $S$ with underlying atoms, see Fig.~\ref{model}.
(Note that $J$-term to the atom just below is irrelevant since it produces a constant contribution to energy, eventually absorbed by normalization factor when probabilities are calculated.)
The presence of the $S$-term, known as the Ehrlich-Schwoebel barrier,\citeup{ehrlich66,schwoebel66} is necessary to account for the so called edge effect when particles do not hop down a terrace edge.

        As we mentioned, the anisotropy will be discussed in terms of the height-height correlation function $G(\vec{s})$ of a planar vector $\vec{s}$, a given distance between the pairs of columns
\begin{equation}
G(\vec{s}) \equiv
\langle h(\vec{r}) \cdot h(\vec{r}+\vec{s})\rangle - \langle h(\vec{r}) \rangle^2,
\label{eq_g}
\end{equation}
where $\langle\ldots\rangle$ denotes spatial average (over all sites $\vec{r}$) and $\langle h(\vec{r}) \rangle$ is the average height of the film.
Such a two-site characteristic is necessary for description of anisotropy.
Note, that $G(0,0)=\sigma^2$, and $\sigma$ is the standard deviation of surface heights.
We also define dimensionless functions $g_x \equiv G(1,0)/G(0,0)$ and $g_y \equiv G(0,1)/G(0,0)$.
For positive values $g_x>0$ (or $g_y>0$), we have terraces-like smooth surfaces in $x$-direction (or $y$-direction).
On the other hand, for negative values of correlation function $g_x<0$ (or $g_y<0$), a spiky and rough structures are observed.
The ellipticity parameter $\varepsilon\equiv g_x-g_y$ may be used as a measure of surface anisotropy.
The ellipticity $\varepsilon$ vanishes for isotropic surface, anisotropy manifests itself as $\varepsilon\ne 0$.

\section{Results of Simulation and Discussion}
\noindent
Simulations were carried out on $500\times 500$ large square lattices with periodic boundary conditions to minimize boundary effects.
The average coverage was ten monolayers (ML).
        As we mentioned earlier, the anisotropic growth could be provoked either by anisotropy in in-plane interaction energy $J_x\neq J_y$ (see Table~\ref{tab_asym} and \ref{tab_antysym}), in binding to underlying layer energy $S_x\neq S_y$ (Table~\ref{tab_ellip_v_s}), and/or by anisotropy in the diffusion barrier $V_x\neq V_y$ when $J\neq 0$ (Table~\ref{tab_ellip_v_j}).

%
%
\begin{table}
\centering
\caption{Ellipticity $\varepsilon$ for $J_x=J$, $J_y=0$ (asymmetric case) for different values of diffusion barrier $V_x=V_y=V$.
$S_x=S_y=0$.}
\label{tab_asym}
\begin{tabular}{r | rrrrrrrrr}
\hline
$V/J$
   &   -5&   -1&  -0.5& -0.25&    0&  0.25&   0.5&   1.0&   5.0\\
\hline
  0& 0.97& 0.61&  0.43&  0.25& 0.00& -0.28& -0.50& -0.71& -0.73\\
  1& 0.93& 0.58&  0.40&  0.23& 0.00& -0.23& -0.39& -0.54& -0.66\\
  2& 0.89& 0.50&  0.31&  0.17& 0.00& -0.16& -0.28& -0.42& -0.61\\
  3& 0.86& 0.37&  0.20&  0.10& 0.00& -0.09& -0.17& -0.31& -0.60\\
  4& 0.80& 0.22&  0.09&  0.04& 0.00& -0.04& -0.09& -0.20& -0.57\\
  5& 0.72& 0.11&  0.03&  0.02& 0.00& -0.02& -0.03& -0.10& -0.54\\
 10& 0.41& 0.00&  0.00&  0.00& 0.00&  0.00&  0.00&  0.00& -0.45\\
 15& 0.01& 0.00&  0.00&  0.00& 0.00&  0.00&  0.00&  0.00& -0.01\\
 20& 0.00& 0.00&  0.00&  0.00& 0.00&  0.00&  0.00&  0.00&  0.00\\
\hline
\end{tabular}
\end{table}

\begin{figure}
\centering
\includegraphics[width=30mm,angle=-90]{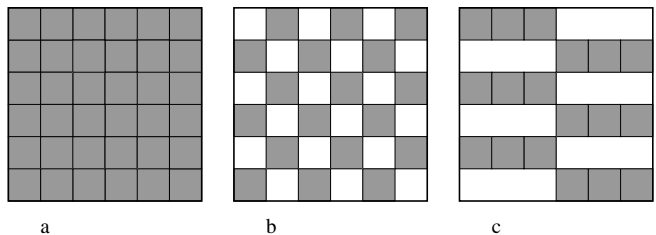}
\caption{Schematic sketch of some typical situation: (a) smooth surface, (b) rough surface, (c) anisotropic surface.}
\label{sketch}
\end{figure}

        For antisymmetric values of the in-plane interaction energy $J_x=-J_y$, the correlation function in the two perpendicular directions is also antisymmetric, $g_x=-g_y$ (see Table~\ref{tab_antysym}).
In such case particles repel each other in one direction while they attract in the perpendicular one.
The surface profile is then similar to the sketch presented in Fig.~\ref{sketch}c.
The anisotropy parameter $\varepsilon$ does not depend on surface average coverage $\langle h \rangle$ (except of the earliest stages of simulation) as presented in Table~\ref{tab_ell_h}.

%
%
\begin{table}
\centering
\caption{Ellipticity $\varepsilon$ for $J_x=-J_y=J$ (antisymmetric case) for different values of diffusion barrier $V_x=V_y=V$.
$S_x=S_y=0$.}
\label{tab_antysym}
\begin{tabular}{r | rrrrrrrrr}
\hline
$V/J$
   &    -5&    -1&  -0.5& -0.25&     0&  0.25&   0.5&     1&     5\\
\hline
  0&  1.58&  1.47&  1.07&  0.57&  0.00& -0.57& -1.07& -1.47& -1.57\\
  1&  1.58&  1.42&  0.97&  0.50&  0.00& -0.51& -0.97& -1.42& -1.57\\
  2&  1.57&  1.28&  0.72&  0.35&  0.00& -0.36& -0.72& -1.28& -1.57\\
  3&  1.57&  0.94&  0.40&  0.18&  0.00& -0.19& -0.40& -0.94& -1.57\\
  4&  1.57&  0.50&  0.18&  0.08&  0.00& -0.08& -0.18& -0.50& -1.56\\
  5&  1.55&  0.22&  0.07&  0.03&  0.00& -0.03& -0.07& -0.22& -1.55\\
 10&  1.28&  0.00&  0.00&  0.00&  0.00&  0.00&  0.00&  0.00& -1.27\\
 15&  0.31&  0.00&  0.00&  0.00&  0.00&  0.00&  0.00&  0.00& -0.31\\
 20&  0.03&  0.00&  0.00&  0.00&  0.00&  0.00&  0.00&  0.00&  0.03\\
\hline
\end{tabular}
\end{table}

%
%
\begin{table}
\centering
\caption{Ellipticity $\varepsilon$ for $S_x=S$, $S_y=0$ for different values of diffusion barrier $V_x=V_y=V$.
$J_x=J_y=0$.}
\label{tab_ellip_v_s}
\begin{tabular}{r | rrrrrrrrr}
\hline
$V/S$
   &   -5&   -2&   -1&-0.75& -0.5&-0.25&     0\\
\hline
  0& 0.80& 0.68& 0.55& 0.47& 0.38& 0.23& 0.00\\
  1& 0.78& 0.67& 0.52& 0.45& 0.36& 0.21& 0.00\\
  2& 0.76& 0.61& 0.45& 0.38& 0.28& 0.16& 0.00\\
  3& 0.75& 0.54& 0.35& 0.26& 0.18& 0.09& 0.00\\
  4& 0.71& 0.43& 0.20& 0.13& 0.08& 0.04& 0.00\\
  5& 0.66& 0.32& 0.09& 0.05& 0.03& 0.02& 0.00\\
 10& 0.37& 0.00& 0.00& 0.00& 0.00& 0.00& 0.00\\
 15& 0.02& 0.00& 0.00& 0.00& 0.00& 0.00& 0.00\\
 20& 0.00& 0.00& 0.00& 0.00& 0.00& 0.00& 0.00\\
\hline
\end{tabular}
\end{table}

%
%
\begin{table}
\centering
\caption{Ellipticity $\varepsilon$ for $V_x=V$, $V_y=0$ for different values of binding energy $J_x=J_y=J$.
$S_x=S_y=0$.}
\label{tab_ellip_v_j}
\begin{tabular}{r | rrrrrrrrrrr}
\hline
$V/J$
   &    -5&    -2&    -1&     0&     1&     2&     5\\
\hline
  0&  0.00&  0.00&  0.00&  0.00&  0.00&  0.00&  0.00\\
  1& -0.02& -0.04& -0.07&  0.00&  0.03&  0.02&  0.00\\
  2& -0.03& -0.09& -0.16&  0.00&  0.06&  0.05&  0.01\\
  3& -0.05& -0.15& -0.29&  0.00&  0.07&  0.03&  0.02\\
  4& -0.06& -0.23& -0.40&  0.00&  0.08&  0.08&  0.03\\
  5& -0.09& -0.34& -0.47&  0.00&  0.08&  0.08&  0.05\\
 10& -0.27& -0.60& -0.53&  0.00&  0.08&  0.09&  0.09\\
 15& -0.51& -0.60& -0.53&  0.00&  0.08&  0.09&  0.08\\
 20& -0.64& -0.61& -0.53&  0.00&  0.08&  0.09&  0.08\\
\hline
\end{tabular}
\end{table}

	The anisotropy parameter $\varepsilon$ depends on the ratio $|J_x-J_y|/|J_x+J_y|$ of the parameters and not on the differences.
For the asymmetric case we found that the ellipticity $\varepsilon$ tends to zero for large values of binding energy although $J_x-J_y\neq 0$ (see Table~\ref{ell_mj} and~\ref{ell_pj}).
The anisotropy vanishes also for relatively large diffusion barrier $V$ (Table~\ref{tab_asym}, \ref{tab_antysym} and \ref{tab_ellip_v_s}).

%
%
\begin{table}
\centering
\caption{Roughness $\sigma$ [ML] and ellipticity $\varepsilon$ for asymmetric anisotropic case ($J_x=J$ and $J_y=J-10$).}
\label{ell_mj}
\begin{tabular}{r rrrrr}
\hline
$J$           & -10.00& -5.00& -2.00& -1.00&  0.00\\
$\sigma$      &   0.95&  0.94&  0.99&  1.07&  1.78\\
$\varepsilon$ &  -0.08& -0.09& -0.16& -0.28& -1.03\\
\hline
\end{tabular}
\end{table}

%
%
\begin{table}
\centering
\caption{Roughness $\sigma$ [ML] and ellipticity $\varepsilon$ for asymmetric anisotropic case ($J_x=J$ and $J_y=J+10$).}
\label{ell_pj}
\begin{tabular}{r rrrrr}
\hline
$J$           &  0.00&  1.00&  2.00&  5.00& 10.00\\
$\sigma$      & 19.04& 20.63& 21.85& 22.55& 22.57\\
$\varepsilon$ &  0.75&  0.31&  0.11&  0.00&  0.00\\
\hline
\end{tabular}
\end{table}

\begin{table}
\centering
\caption{Dependence of dimensionless correlation functions in perpendicular direction ($g_x, g_y$) for antisymmetric case $J_x=-J_y=-2.0$ for different film average heights $\langle h \rangle$ [ML].}
\label{tab_ell_h}
\begin{tabular}{r rrrrrrrr}
\hline
$\langle h \rangle$&    1&    5&   10&   25&   50&   75&  100&  500\\
$g_x$              & 0.54& 0.75& 0.79& 0.80& 0.80& 0.81& 0.80& 0.80\\
$g_y$              &-0.45&-0.73&-0.80&-0.81&-0.82&-0.83&-0.83&-0.84\\
\hline
\end{tabular}
\end{table}
%
%
\begin{table}
\centering
\caption{Correlations expressed in $G(0,0)$ units for anisotropic case ($J_x=-J_y=-2.0$).}
\label{gxgyaniso}
\begin{tabular}{r rrrrr}
\hline
     $n$&     1&    2&     3&    4&     5\\
$G(0,n)$& -0.80& 0.49& -0.24& 0.10& -0.03\\
$G(n,0)$&  0.79& 0.53&  0.31& 0.15&  0.06\\
$G(n,n)$& -0.66& 0.29& -0.09& 0.01&  0.00\\
\hline
\end{tabular}
\end{table}

	Anisotropic growth of the surfaces follows anisotropy in either the atom pairs binding energy, or in different diffusion barriers and mobility of atoms in different directions.
Surface roughness, which is usually characterized by the width of the surface, is not a suitable choice for anisotropy discussion as a single site property only.
Instead we need a two-site properties such as the height-height correlation function $G(n,m)$ for columns separated by a vector $(n,m)$, or ellipticity parameter $\varepsilon$.
The main results of the simulations may be summarized as follows:

	Firstly, for the isotropic case the correlation function $G(1,0)$ or $G(0,1)$ itself reflects the roughness.
For random deposition $G(1,0)=G(0,1)=0$.
For flatter surfaces we claim positive $G(1,0)$ and $G(0,1)$.
This is so since any two adjacent sites are expected to have the same heights and so the deviation from the average film thickness are of the same sign resulting in the positive contribution to the correlation.
And on the contrary, rough surfaces show negative $G(1,0)$ and $G(0,1)$.
This, however, does not yield information on possible anisotropy of the surface when $G(1,0) \ne G(0,1)$.
Then anisotropy parameter $\varepsilon$ properly normalized to $G(0,0)$ is used.

	In the limiting case of large diffusion barriers the migration stops and again we restore the random deposition model with its consequences, correlations $G(\vec{s}\ne\vec{0})=0$ and anisotropy $\varepsilon=0$.
However, if migration takes place then $G(\vec{s}\ne\vec{0})\ne 0$ and we may expect that only for distant sites $G(n,m)\to 0$, when $n\to\infty$ and/or $m\to\infty$.
In fact, we often observed a damped character of $G(n,0)$ dependence on the distance $n$ between atoms in the $x$-direction (or similarly $G(0,n)$ in the $y$-direction).
The roughly exponential character of the decaying $G(n,0)$ and $G(0,n)$ on $n$ may be observed for the direction which shows a smooth variations in column width (negative $J$, small $\sigma$).
The decrease of $G(n,0)$ and $G(0,n)$ on $n$ may have an oscillatory character changing the sign of $G(n,0)$ or $G(0,n)$ if rough and spiky structure (positive $J$, large $\sigma$) is observed.
Therefore anisotropic case $J_x=-J_y$ produces different $G(n,0)$ and $G(0,n)$ on $n$ dependence in $x$- and $y$-directions, exponential decrease along $x$-axis for atom pairs correlation $G(n,0)$, and damped oscillatory dependence of $G(0,n)$ in $y$-direction, as it is shown in Table~\ref{gxgyaniso}.

	The close-distance correlations $G(1,0)$, $G(0,1)$ and $G(1,1)$ may also be helpful to distinguish between different classes of surface morphology.
As we mentioned earlier, $G(1,0)<0$ ($G(0,1)<0$) indicates rough surface, $G(1,0)>0$ ($G(0,1)>0$) indicates flat surface.
Therefore, we expect for an isotropic flat surface as in Fig.~\ref{sketch}a $G(1,0)=G(0,1)>0$ and $G(1,1)>0$.
For a chessboard-like surface, Fig.~\ref{sketch}b, we get $G(1,0)=G(0,1)<0$ and again $G(1,1)>0$.
Thus $G(1,1)$ cannot distinguish between isotropic smooth (Fig.~\ref{sketch}a) and isotropic rough (Fig.~\ref{sketch}b) case. 
However, for anisotropic case, Fig.~\ref{sketch}c, of a terrace-like structure we get $G(1,0)>0$, $G(0,1)<0$ and $G(1,1)<0$. 
We conclude that combination of all $G(1,0)$, $G(0,1)$ and $G(1,1)$ information is indicative of the type of the film morphologies, its roughness and anisotropy.

	The anisotropy in surface morphology, particles migrations and accommodation coefficients, as well as surface magnetic properties, were verified experimentally.
For instance, Mo {\it et al} works\cite{mo89a}$^-$\cite{mo92} show a highly anisotropic diffusion of Si and Ge atoms on a Si(001) surface, with the easy diffusion direction along the dimer rows.
Surface migration of atoms is at least one thousand times faster along the substrate dimer rows than in the direction perpendicular to them.\citeup{mo91b,mo91c}
It could be explained theoretically by assuming the difference in energy activation for atom diffusion $\Delta E=E_{\parallel}-E_{\perp}\approx 0.4$~eV.\cite{mo91c}

\nonumsection{Acknowledgments}
\noindent
Main calculations were carried out in ACC-CYFRONET-AGH.
This work and machine time in ACC-CYFRONET-AGH are supported by Polish Committee for Scientific Research (KBN) with grants no. 8~T11F~02616 and KBN/\-S2000/\-AGH/\-069/\-1998, respectively.

\nonumsection{References}


\end{document}